\documentclass[aip,reprint,twocolumn]{revtex4-1}
\pdfoutput=1
\usepackage{amsmath}
\usepackage{graphicx,color}
\usepackage{booktabs}
\usepackage{datetime}
\definecolor{navyblue}{rgb}{0.0,0.0,1}
\usepackage{xcolor}
\usepackage{epstopdf}
\usepackage{wrapfig}
\usepackage{ragged2e}
\usepackage{epsfig}
\usepackage{slashed}
\begin{document}
\title{Collision induced amplification of wakes in streaming plasmas}

%\author{a b c}

\author{Sita Sundar, Hanno K{\"a}hlert, Jan-Philip Joost, Patrick Ludwig, and Michael Bonitz
 }
%\author{SS, MKV, AGC, and AA}

\affiliation{Institut f\"ur Theoretische Physik und Astrophysik, Christian-Albrechts-Universit{\"a}t zu Kiel, Leibnizstrasse 15, Kiel 24098, Germany}
\begin{abstract}
This work examines the formation of wake fields caused by ions streaming around a charged dust particle, using three-dimensional particle-in-cell (PIC) simulations with charge-neutral collisions included. The influence of an external driving electric field, which leads to a non-Maxwellian distribution of ions, is investigated in detail. The wake features formed for non-Maxwellian ions exhibit significant deviations from those observed within the model of a shifted Maxwellian distribution. The dependence of the peak amplitude and position of the wake potential upon the degree of collisionality is analyzed for a wide range of streaming velocities (Mach numbers). In contrast to a shifted Maxwellian distribution of ions, the drift-driven non-Maxwellian distribution exhibits an increase of the wake amplitude of the first attractive peak with increase in collisionality for high streaming velocities. At very low Mach numbers, collision-induced amplification is observed for Maxwellian as well as non-Maxwellian distributions. 
\end{abstract}

\maketitle
%upto here 
\section{Introduction}
Complex (or dusty) plasmas constitute a well-studied~\cite{Morfill:RMP2009,Melzer:WVV2008,Bonitz:Book2010,bonitz_2010_complex} interdisciplinary research area. Due to the presence of micron-sized, highly charged dust particles in addition to electrons, ions and neutrals, these plasmas are endowed with features like self-organization and crystal-structures~\cite{Chu:PRL1994}, low-frequency waves (dust acoustic modes)~\cite{barkan:POP1995}, or the formation of dust-free structures (voids) in micro-gravity experiments~\cite{Kretschmer:PRE2005}. Complex plasmas have close connections with astrophysics (e.g., dust in planetary rings or cometary tails), low-temperature gas discharge operation (dust inserted into or grown in-situ in gas discharges)~\citep{Barbosa:JPD2016}, fusion related research (dust generated in plasma-surface interactions)~\citep{Winter:PPCF2004, Sharpe2002153, Ratyn:PPCF2008}, warm-dense matter (strong coupling physics)~\cite{graziani-book,zhandos_pop_15} as well as the broad regime of material sciences (e.g., defects or wave propagation in dust crystals)~\cite{Hartmann:PRL2009,Hartmann:PRL2010}.

When a dust particle is inserted into a low-temperature plasma, its potential is screened by the electrons and ions. In the sheath region of gas discharges, where the dust typically levitates, strong electric fields and significant ion flows occur. It has been established~\cite{Lampe:POP2000,Winske:IEEE2001} that the deflection of streaming ions in the electric field of the dust particle leads to quasi-periodic waveforms downstream of the grain. In the linear response (LR) formalism, e.g.~\cite{Lampe:POP2000,Ludwig:NJP2012,block:CTPP2012}, the wake potential is linearly proportional to the grain charge, $Q_d$. Winske \textit{et al}.~\cite{Winske:IEEE2001}, however, found a nonlinear relationship between charge and wake potential utilizing one- and two-dimensional PIC-simulations. Hutchinson et al.~\citep{Hutch:POP2011} also reported the nonlinear suppression of the peak potential in the oscillatory wake structure and a reasonable agreement with linear response calculations in the linear regime.

Kinetic models that study the effect of streaming ions on a system of dust particles often adopt a distribution function that is a shifted Maxwellian (MW). However, Lampe et al.~\cite{Lampe:POP2012} have calculated the steady state distribution function $f(\vec v)$ for ions subject to ion-neutral charge-exchange collisions in the presence of an electric field, using Monte-Carlo simulations, and obtained a non-Maxwellian ditsribution~\cite{Lampe:POP2012}. Their result is quite distinct from the shifted Maxwellian distribution. The physical characteristics of the non-Maxwellian distribution are due to the presence of collisions and the force field that gives rise to the ion drift. This modified distribution has been shown to influence the ion drag on the grain~\cite{ivlev2005, Patacchini:PRL2008}, the ion-dust streaming instability~\cite{Hanno:POP2015}, and also affects profoundly the physics of wake formation~\cite{Hutch:POP2013, Kompaneets:PRL2016, KompaneetsPRE}. While Maxwellian ions give rise to several potential minima and maxima downstream of the grain~\cite{Hutch:PRE2012}, non-Maxwellian ions ususally exhibit only one pronounced maximum~\cite{Hutch:POP2013}. However, a detailed and systematic numerical analysis of non-Maxwellian effects on the nonlinear wake potential is still missing.

Therefore, in this work, we provide such an analysis. In particular, we present a wide-ranging numerical exploration of the role of charge-exchange collisions and the ion distribution function on the wake formation around a grain in a uniform streaming plasma using the three-dimensional Particle-in-Cell (PIC) simulation code COPTIC~\cite{Hutch:POP2011}. In contrast to linear response calculations~\cite{Kompaneets:PRL2016, KompaneetsPRE}, this allows us to fully account for nonlinear effects, which are expected to play an important role in the close vicinity of the grain. We calculate the wake potential for various conditions and study the effect of the distribution function on the wake formation. One of the most interesting results is that {\em collisions do not necessarily weaken or destroy wake effects}, as is usually expected. In contrast, we report that {\em charge-exchange collisions may even enhance wake effects}, and we present an explanation for this observation.

The outline of the paper is as follows. In Sec.~\ref{sec:distribution}, we present a brief description of the distribution functions and their physical interpretation. This is followed by details of the three-dimensional particle-in-cell simulations (COPTIC), in Sec.~\ref{sec:PIC}. In Sec.~\ref{sec:wake}, we discuss the topology of the potential, the influence of the ion distribution function (Maxwellian vs. non-Maxwellian), and present systematic results (comparison of peak heights, peak positions) for a broad range of plasma parameters, including a wide range of Mach numbers and collision frequencies.  Finally, we present a summary and our conclusions in Sec.~\ref{sec:conclusion}.

\section{Ion Distribution Function}\label{sec:distribution}
The simple choice of a {\it shifted Maxwellian distribution} is often a good description of streaming effects without external fields and without any significant effect of collisions. However, in the sheath region of discharges, the ions undergo collisions with the neutrals and experience a strong acceleration towards the electrodes by the sheath electric field. This has been shown to affect their distribution function considerably~\cite{Lampe:POP2000}. Even the inclusion of a small degree of collisionality and a (uniform) external field (mean electric field) affect the distribution. The main difference compared to a shifted Maxwellian with the same drift velocity is that the selfconsistently calculated distribution retains a considerable fraction of low-velocity ions. These distributions have been named {\it drift driven distributions} (non-Maxwellian distributions). The nomenclature of {\it shifted Maxwellian distribution} and {\it drift-driven distribution} that has been adopted in Ref.~\cite{Hutch:POP2013} will be used throughout this paper.

In the picture of the {\it shifted Maxwellian distribution}, the drift velocity of the neutral particles is the same as that of the ions. There is no external electric field explicitly taken into account, and the ion flow is driven by the neutral flow. As a result, the shifted ion distribution and the neutral distribution are identical, which also implies that their temperatures are equal, $T_i=T_n$. Hence, after a collision, the new ion has a net (neutral) flow velocity. In contrast, in the case of the {\it drift driven distribution}, the drift velocity of neutrals is zero, and the ion flow is driven by the external electric field. Thus, ion-neutral charge-exchange collisions give rise to ions with zero drift velocity upon collision, see also Ref.~\cite{Hutch:POP2013}.

\begin{figure}
\includegraphics[width=0.5\textwidth]{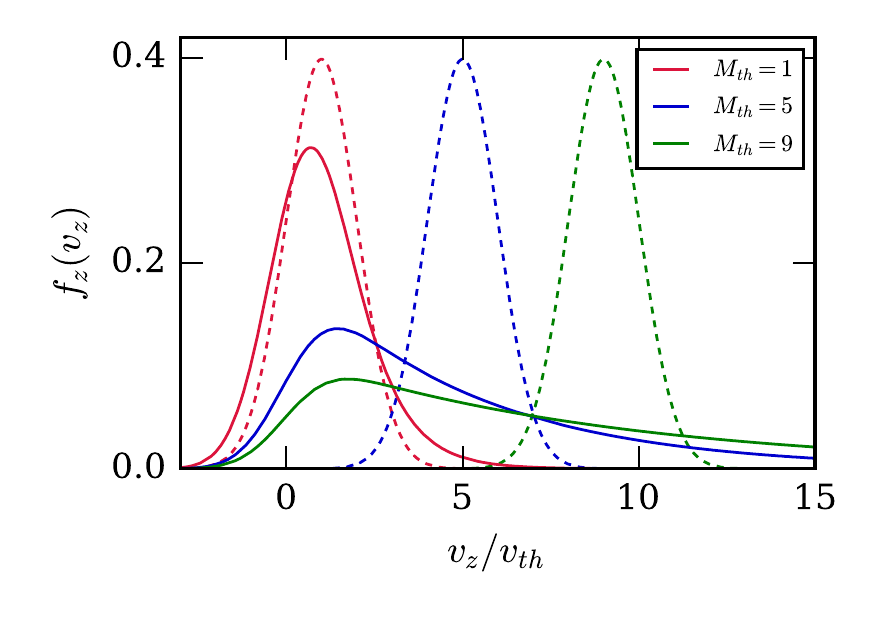}
\caption{Shift and drift velocity distribution function along the external field plotted versus velocity (normalized to the neutral thermal velocity) for three different thermal Mach numbers. The full lines depict the drift-driven distribution function [Eq.~\eqref{eqn:drift}] while the dashed lines represent the corresponding shifted Maxwellian distributions.}
\label{fig:figure1}
\end{figure}

For cases where the contribution of ion-ion collisions is smaller than that of ion-neutral charge exchange collision (ig. for low degree of ionization), the drift solution of the Boltzmann equation with a BGK-collision term (relaxation time approximation with constant ion-neutral collision frequency $\nu_{in}$) in the presence of a uniform external driving field $E_0$ can be delineated analytically as~\citep{Lampe:POP2012}
\begin{align}\label{eqn:drift}
f_z(u_z)&= \frac{1}{2M_\text{th}}\exp \left(\frac{1-2M_\text{th}u_z}{2M_\text{th}^2}\right) \times \nonumber\\
&\phantom{=}\left[1+\text{erf}\left(\frac{M_\text{th} u_z-1}{\sqrt{2}M_\text{th}}\right)\right].
\end{align}
Here, $u_z=v_z/v_\text{th}$ denotes the velocity in the streaming direction (normalized to the thermal velocity of neutrals, $v_\text{th}$), and $M_\text{th}=v_d/v_\text{th}$ the thermal Mach number, where $v_d=q E_0/(m\nu_\text{in})$ is the drift velocity (ion charge $q$). To illustrate the two kinds of ion distributions, we present a comparison of the shifted Maxwellian distribution and the drift-driven distribution in Fig.~\ref{fig:figure1}. One can clearly see that the difference in the two distributions increases for higher Mach numbers.

\section{Particle-in-Cell simulations}\label{sec:PIC}
For our simulations, we have used the three-dimensional Cartesian mesh, oblique boundary, particles and thermals in cell (COPTIC) code~\cite{Hutch:POP2011}. COPTIC is a hybrid particle-in-cell code in a sense that electron dynamics are governed by the Boltzmann description, $n_e = n_{e \infty}\exp(e\phi/T_e)$, whereas ion dynamics are considered in six-dimensional phase space in the presence of the self-consistent electric field and an optional external field. Length scale, velocity and other normalizations are adopted from the paper by Hutchinson et al.~\cite{Hutch:POP2011}. In this work, we have considered only point-charge grains using the particle-particle-particle-mesh (PPPM) scheme~\cite{Hockney:Book1988}. The magnitude of the point charge is specified by an input value, and it equals the Coulomb potential (in normalized units) that would exist at the object's radius if the charge were unshielded.~\citep{Hutch:POP2011} The potential for a point charge (dust grain) is represented by an analytic part that is non-zero up to a specified radius. Outside that radial distance, the analytic part is zero.

Collisions are incorporated according to a Poisson statistical distribution with fixed velocity-independent collision frequency, i.e., the same assumption as used in the derivation of Eq.~\eqref{eqn:drift} (BGK-type collisions). Charge-exchange collisions of ions are performed by exchanging the velocity of the colliding ion with the velocity of a neutral chosen randomly from the neutral velocity distribution \cite{Hutch:POP2011}.

\section{Analysis of the wake potential}\label{sec:wake}
We have performed a systematic study of the wake potential as a function of both, the Mach number and the collision frequency. The electron-ion temperature ratio was fixed at $T_e/T_i =100$. To delineate the differences in the wake potential for the Maxwellian and non-Maxwellian distribution, we performed simulations for parameters in the range $M=v_d/c_s=\sqrt{T_i/T_e}\, M_\text{th}$ = 0.2, 0.5, 1, 1.5 and $\nu/\omega_{pi}$ = 0.001, 0.01, 0.1, 0.5, where $\omega_{pi}$ is the ion plasma frequency and $c_s=\sqrt{k_B T_e/m_i}$ the ion sound speed.

\subsection{Scaling with grain charge}
\begin{figure}
%%%%%%%%%%%%%%%%%%%%
\hspace{-0.8cm}
\includegraphics[width=0.53\textwidth, trim = 0cm 6cm 7cm 0cm, clip =true,  angle=0]{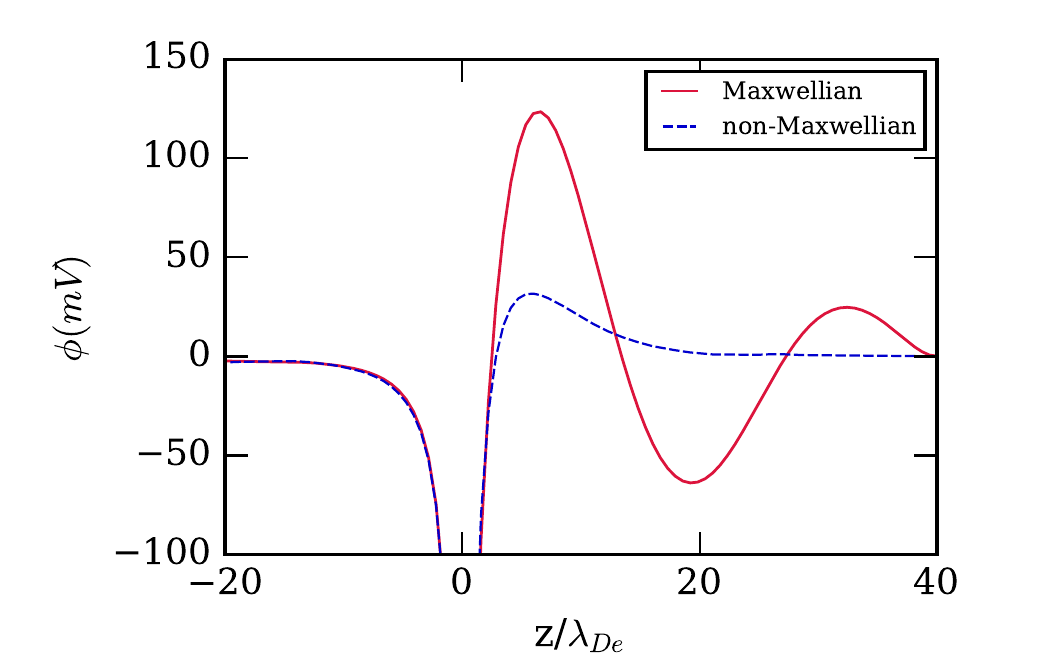}
%%%%%%%%%%%%%%%%%%%%%%%%%%%%
\caption{Wake potential along the streaming axis for $M=0.8$ ($M_{th}=8$) and $\nu/\omega_{pi}=0.01$. The solid (dashed) line corresponds to the shifted Maxwellian (non-Maxwellian) distribution.}
%%%%%%%%%%%%%%%%%%%%%%%
\label{fig:wakestreamingdir}
\end{figure}

Typical examples for the wake potential, as obtained from COPTIC, are shown in Fig.~\ref{fig:wakestreamingdir}. The potential shows one or several oscillations along the streaming direction, depending on the plasma conditions. By definition, in the linear regime, the potential $\Phi(\vec r)$ is linearly proportional to the grain charge. In order to investigate the scaling behavior with PIC simulations, we varied the input grain charge (which is defined, in COPTIC, in terms of the grain potential) and calculated the height of the first potential peak.
%%%%%%%%%%%%%%%%%%%%

\begin{figure}
\includegraphics[width=0.5\textwidth]{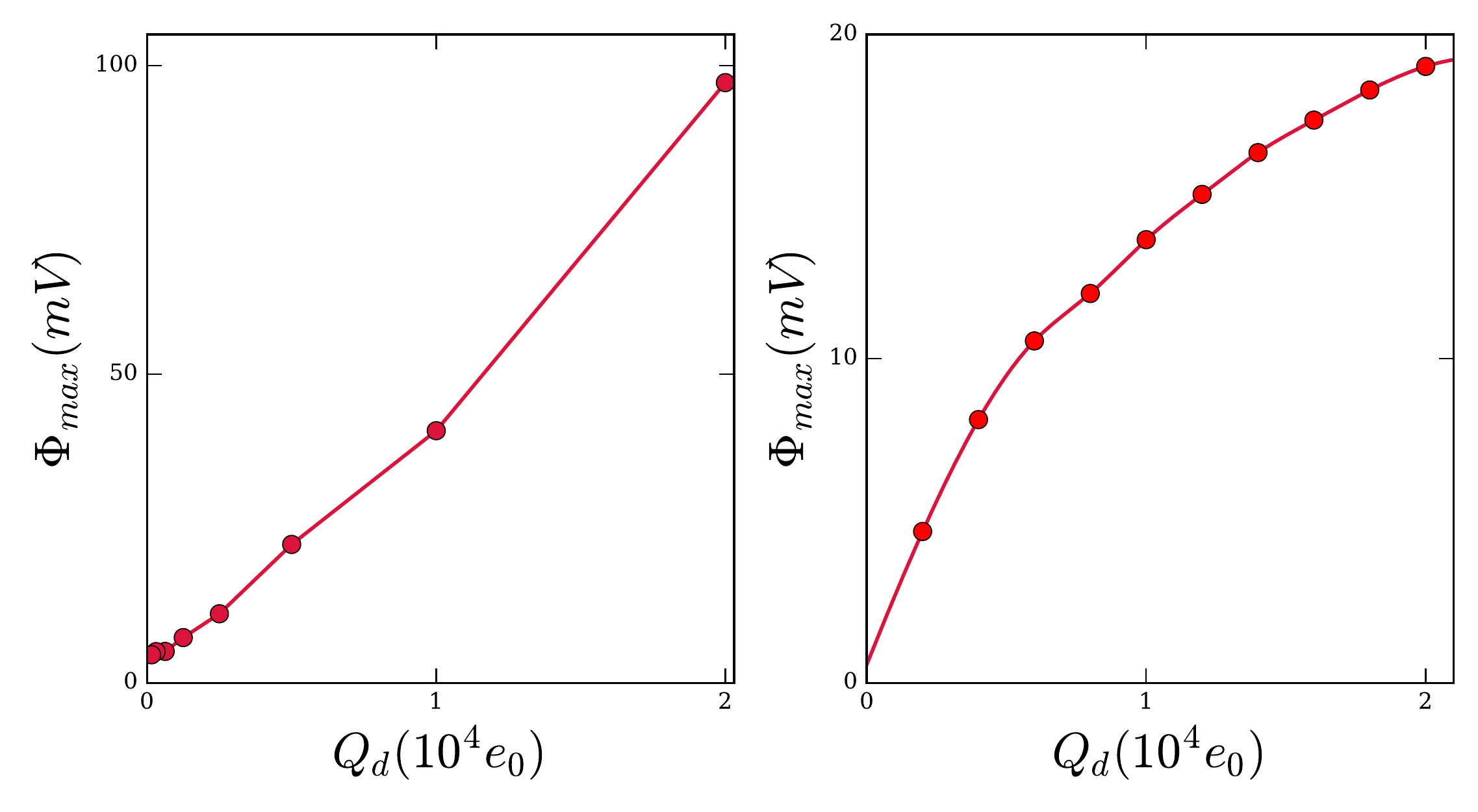}
\hspace*{1.1cm}(a) \hspace*{3.6cm} (b)
\caption{Variation of the maximum of the wake potential with object charge for (a) the shifted Maxwellian distribution and (b) the non-Maxwellian field-driven distribution at $M=1$ ($M_{th}=10$) and $\nu/\omega_{pi}=0.1$.}
\label{fig:figure3}
\end{figure}

The trend of the wake peak height with grain charge is shown in Fig.~\ref{fig:figure3}. It is evident that the two distributions do not obey the same scaling of the peak height with the grain charge. The shifted Maxwellian distribution exhibits an almost linear behavior, up to a maximum charge of $\sim 10^4$ electrons on the grain, while for the same charge on the grain the scaling for the non-Maxwellian distribution is non-linear. Note also that the amplitude of the peak, for the shifted Maxwellian distribution, is significantly higher than for the drift-driven non-Maxwellian case. This is due to the fact that the bunching of ions behind the grain is stronger in the shift case than in the drift case. Even in the nearly linear regime, the effective charge of the grain attracting streaming ions is stronger in the shift case than in the drift case, leading to a higher amplitude wake peak than for the shifted Maxwellian case.

Nonlinear effects should be particularly important for slow ions whose kinetic energy is insufficient to escape from the potential well created by the dust grain. From Fig.~\ref{fig:figure1} it is evident that, in the Maxwellian case, the number of low-velocity ions is very low while in the non-Maxwellian distribution an appreciable fraction of low velocity ions remains even for $M_{th}\approx 10$. This gives an indication as to why the scaling of the peak height with the grain charge does not show a linear behavior, in contrast to the Maxwellian case.

\subsection{Influence of collisionality}
\begin{figure*}
\includegraphics[width=0.8\textwidth,  trim = 1cm 12.4cm 0cm 3cm, clip =true, angle=0]{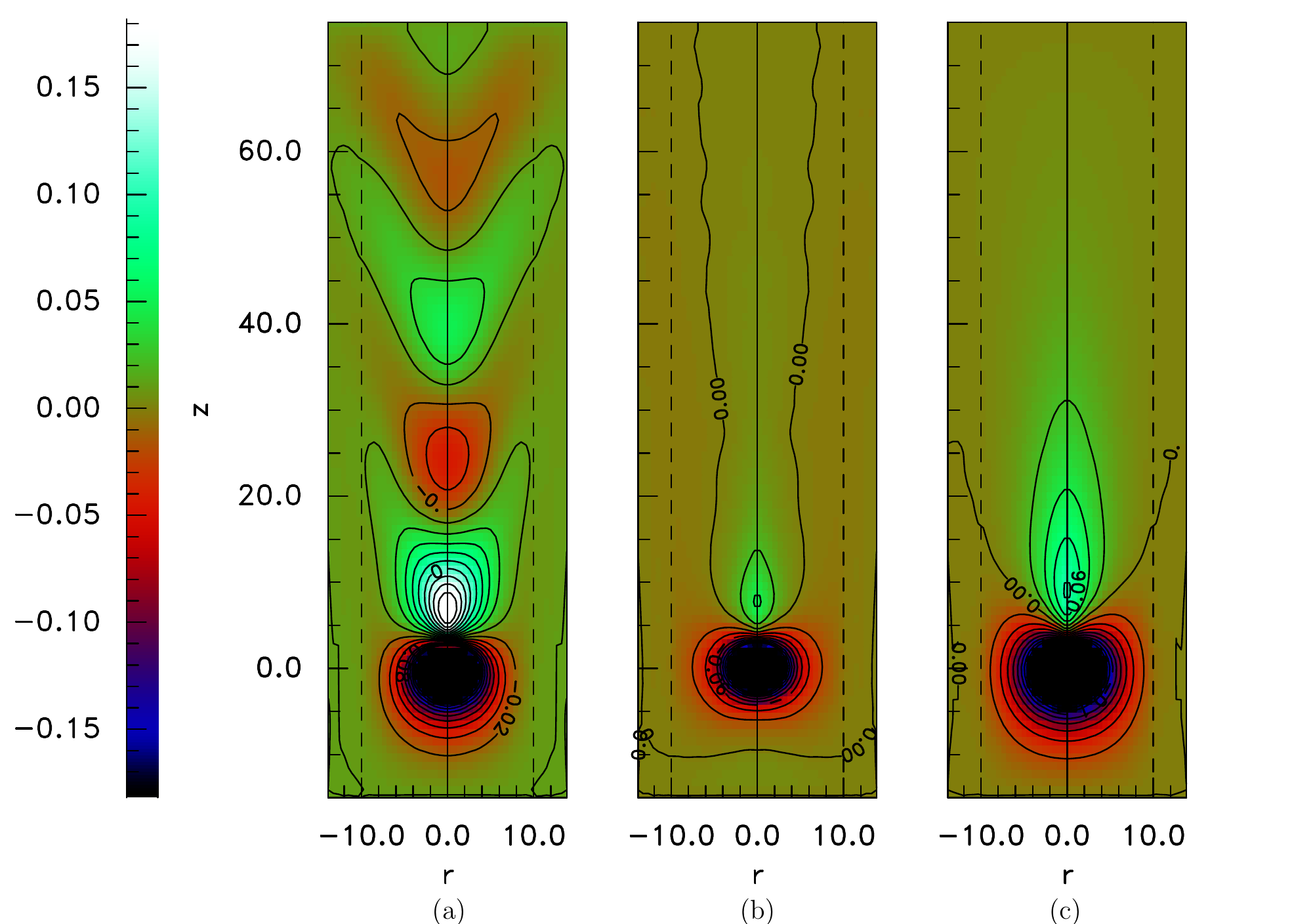}
\caption{Wake potential for (a) the shifted Maxwellian distribution (almost collisionless, $\nu/\omega_{pi}=0.001$); (b) the field-driven non-Maxwellian distribution ($\nu/\omega_{pi}=0.001$) and (c) the shifted Maxwellian with collisions ($\nu/\omega_{pi}=0.5$). The Mach number is identical ($M=1$) in all cases. The grain charge is $2\times 10^4 e_0$.}
\label{fig:figure4}
\end{figure*}
Figure~\ref{fig:figure4} shows the qualitatively similar physical roles played by collisional and collisionless damping for the topology of the potential. Maxwellian ions with low ion-neutral damping lead to pronounced oscillations of the wake potential in the streaming direction, which is well known from linear response calculations~\cite{Lampe:POP2000, Ludwig:NJP2012} and PIC simulations~\citep{Miloch:PPCF2010, Hutch:POP2011,Ludwig:NJP2012, Miloch:POP2010}. However, these oscillations are almost completely absent in the case of non-Maxwellian ions~\cite{Hutch:POP2013}. As can be seen from Fig.~\ref{fig:figure4}, this effect on the wake potential can be closely mimicked in the Maxwellian case by increasing the ion-neutral collision frequency. Even though the damping mechanisms are very different, the eventual effect on the wake oscillations is the same and clearly visible.
\begin{figure*}
\includegraphics[scale=0.8, trim = 0cm 18cm 0cm 3cm, clip =true, angle=0]{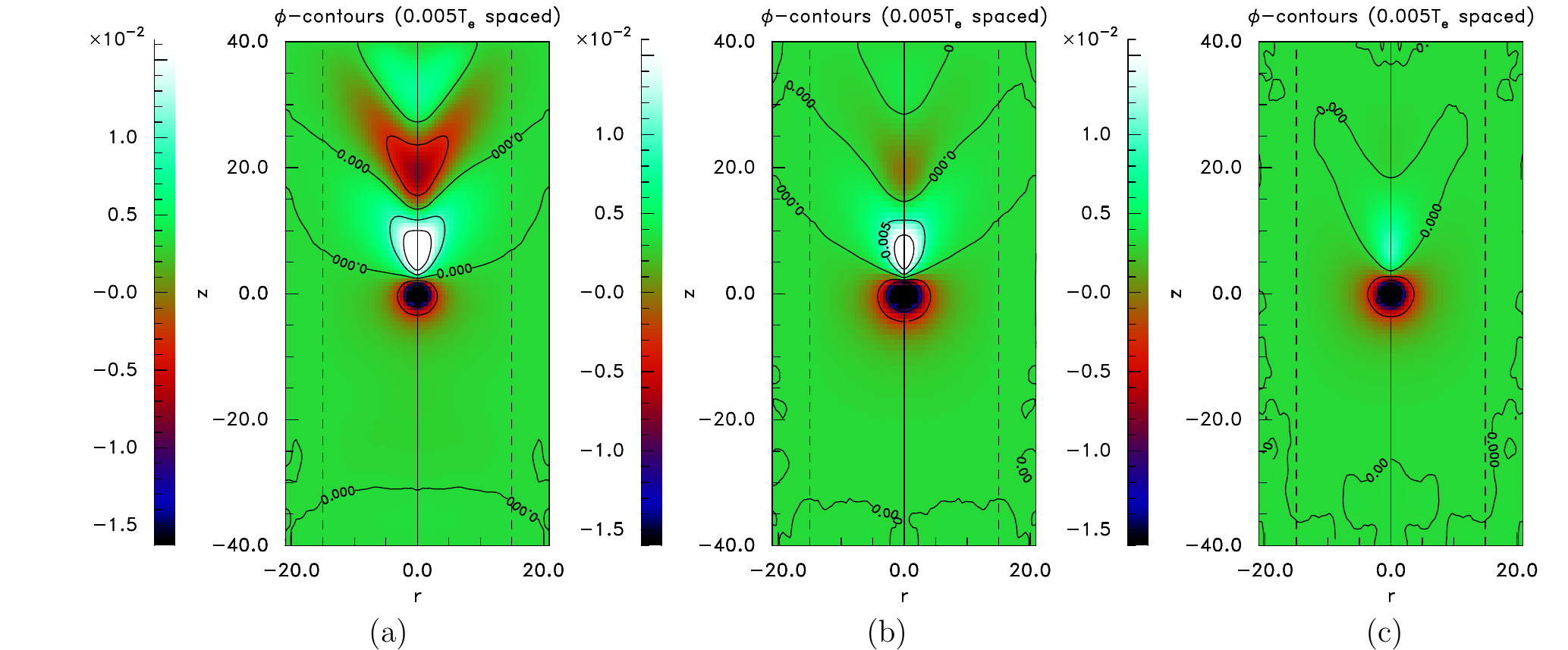} 
\caption{Contour plots of the grain potential $\Phi(\vec r)$ for various collision frequencies: (a) $\nu/\omega_{pi}=0.01$; (b) $\nu/\omega_{pi}=0.1$, and (c) $\nu/\omega_{pi}=0.2$ with streaming velocity $M=0.8$ for the shifted Maxwellian case. The grain charge is $2\times 10^4 e_0$.  }
\label{fig:figure5}
\end{figure*}
\begin{figure*}
\includegraphics[scale=0.8, trim = 0cm 18cm 0cm 3cm, clip =true, angle=0]{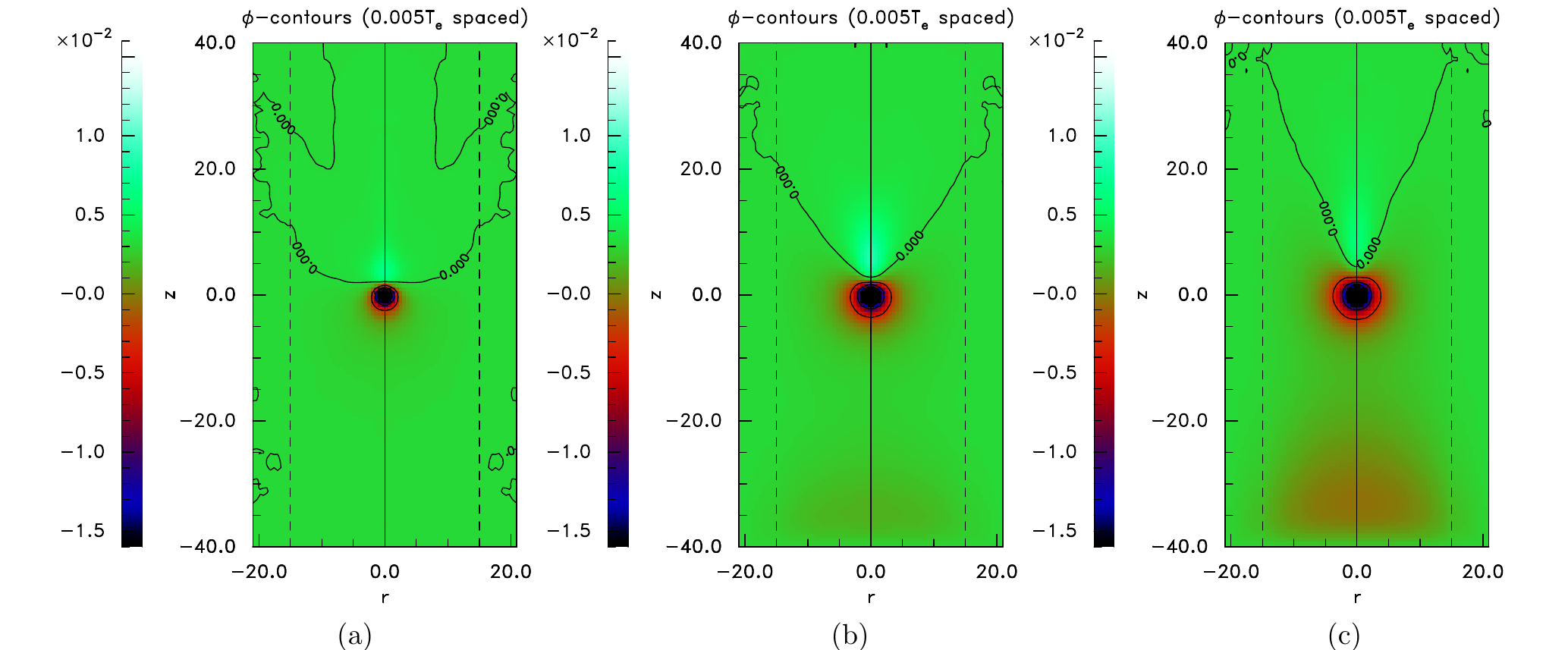} 
\caption{Contour plots of the grain potential $\Phi(\vec r)$ for various collision frequencies: (a) $\nu/\omega_{pi}=0.01$; (b) $\nu/\omega_{pi}=0.1$, and (c) $\nu/\omega_{pi}=0.2$ with streaming velocity $M=0.8$ for the non-Maxwellian case. The grain charge is $2\times 10^4 e_0$. }
\label{fig:figure6}
\end{figure*}

We now investigate the influence of ion-neutral collisions in greater detail. In Fig.~\ref{fig:figure5}, we present wake potential contours, for various collision frequencies, for the shifted Maxwellian distribution. Similar to observations made in linear response calculations~\cite{Ludwig:NJP2012}, here we also observe that the role of collisions is to reduce and damp the wake oscillations behind the grain, for moderate to high Mach numbers~\cite{Hutch:PRE2012}. Figure~\ref{fig:figure6} depicts the same results but now for the non-Maxwellian drift-driven distribution. As observed by Hutchinson \textit{et al.}~\cite{Hutch:POP2013} and in Fig.~\ref{fig:figure4}, the wakes are strongly damped, and only one positive peak behind the grain remains. Unlike in the shifted Maxwellian case, there are no trailing oscillations behind the grain.

\begin{figure*}
\includegraphics[scale=.7, trim = 0.4cm 0cm 0cm 0cm, clip =true, angle=0]{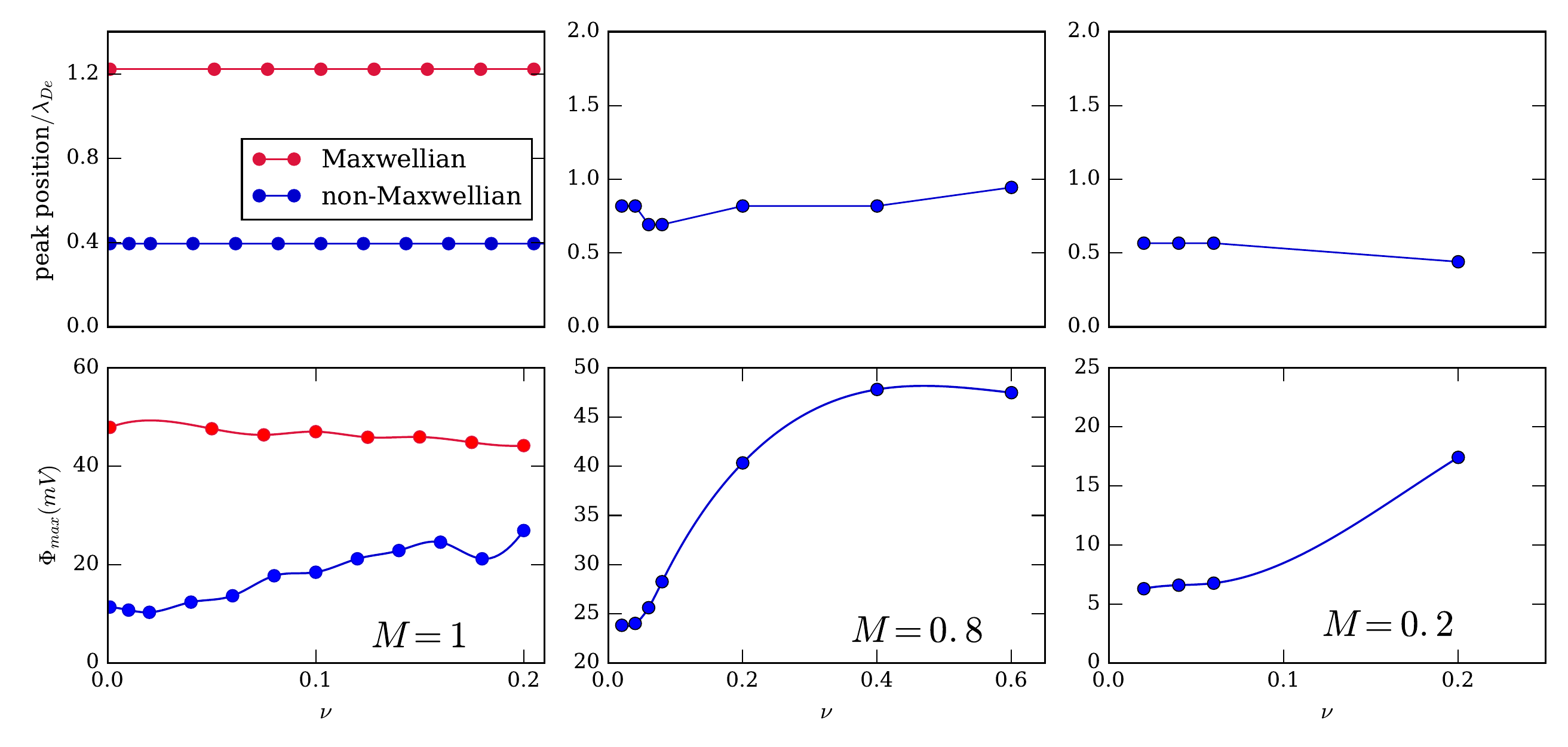}
\caption{Peak position (top row) and peak amplitudes (bottom row) of the wake potential as a function of collision frequency for shifted Maxwellian and non-Maxwellian drift-driven distributions  for $T_e/T_i = 100$ and $M=1, 0.8, 0.2$. }
\label{fig:figure7}
\end{figure*}

At higher collisionality, there is an appreciable fraction of slow ions which eventually lose their energy due to collisions. These slow moving ions cannot escape the potential well and become trapped. The density of these trapped ions builds up over time, and they contribute to the enhancement of the peak amplitude with increased collisionality~\cite{Lampe:POP2003}. A similar discussion with regard to the calculation of drag forces on the grain can be found in Ref.~\cite{Hutch:POP2013}.
\begin{figure}
\includegraphics[scale=0.7, trim = 0.4cm 0cm 0cm 0cm, clip =true, angle=0]{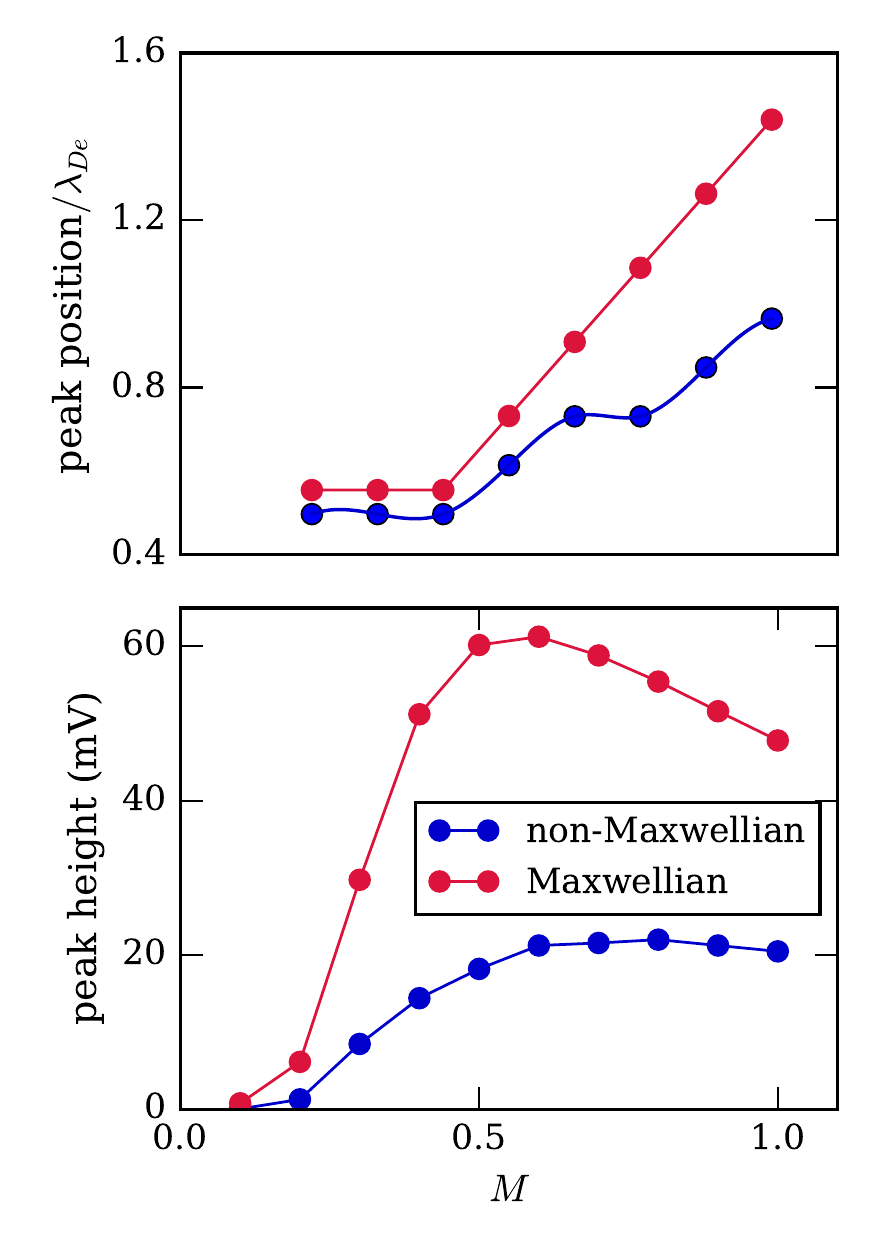}
\caption{Peak position (top) and peak amplitude (bottom) of the wake potential for $T_e/T_i = 100$ as a function of Mach number, for a shifted Maxwellian and a non-Maxwellian drift-driven distribution, for the collision frequency $0.1$.}
\label{fig:figure8}
\end{figure}

The variations of the maximum peak height and the peak position with collisionality, for Maxwellian as well as non-Maxwellian ions, are shown in Fig.~\ref{fig:figure7}. For $M=1$, the peak positions (top left figure) remain constant (within the grid resolution) over the entire range of collision frequencies considered here, for both kinds of distributions. 
For a non-Maxwellian distribution, the peak is substantially closer to the dust grain, compared to the shifted Maxwellian case. Consider now the peak height (bottom left figure). While an increase in collisionality from the collisionless case to $\nu/\omega_{pi}=0.2$ only slightly decreases the height, for the shift distribution, we find that the amplitude for the drift distribution increases significantly, by a factor $\sim 3$. 

In Figure~\ref{fig:figure7} (middle column),  we  present the variations of the maximum peak height and the peak position with collisionality, for the lower flow velocity, $M=0.8$. As in the case $M=1$ we observe collision-induced amplification for small collisionality, $\nu \lesssim 0.2$. When the  collisionality increases further, the peak amplitude exhibits saturation around $\nu \sim 0.4$. Finally, we consider the regime of slow streaming velocity, $M=0.2$, cf. Fig.~\ref{fig:figure7} (right column). Collision-induced amplification is noticed here as well for the non-Maxwellian distribution with a significant increase by a factor $3$ from $\nu=0.01$ to $\nu \sim 0.2$. For larger $\nu$ the increase is less pronounced.
Interestingly, the peak position is slightly reduced with increase of collisionality. The behavior is different for the shifted Maxwellian case. We observe a decrease of the peak height (as in the case $M=1$) for Mach number larger than $0.4$, whereas for smaller values the peak height increases. These results are in agreement with the data presented by Hutchinson~\citep{Hutch:PRE2012} (see, in particular, figure 2 in~\cite{Hutch:PRE2012}) and are not reproduced here.

This peculiar behavior arises from the very nature of the two distributions~\cite{Hutch:POP2013}. In the shifted Maxwellian case, after an ion-neutral charge-exchange collision, the new ion is born with an average velocity equal to the flow velocity. For higher Mach numbers, irrespective of collisionality, fast ions have enough energy to overcome the potential barrier and reduce the ion shielding near the grain, which eventually results in an enhanced shielding length. This is why we do not observe any collision induced amplification of the wake for higher flow velocities (exceeding $M=0.5$, not shown) in the shifted Maxwellian case. On the other hand, in the drift case, an ion that has undergone a collision, is ``reborn'' with zero average velocity. In order to leave the potential well, the ion must first gain a sufficient amount of kinetic energy. At subsonic flow velocities, the behavior is similar to the shifted Maxwellian case. Here also, we see collision-induced amplification of the wake. However, at higher flow velocities, the features are altogether different. At higher flow speeds, the drift-driven non-Maxwellian distribution contains a large fraction of streaming ions in the subsonic range. These subsonic flow ions contribute in shielding and enhancement of the wake amplitude with collision. Thus, collisions lead to an amplification of the maximum wake amplitude, even for high Mach numbers, for a non-Maxwellian distribution.

\begin{figure*}
\includegraphics[scale=.8, trim = 0cm 0cm 0cm 0cm, clip =true, angle=0]{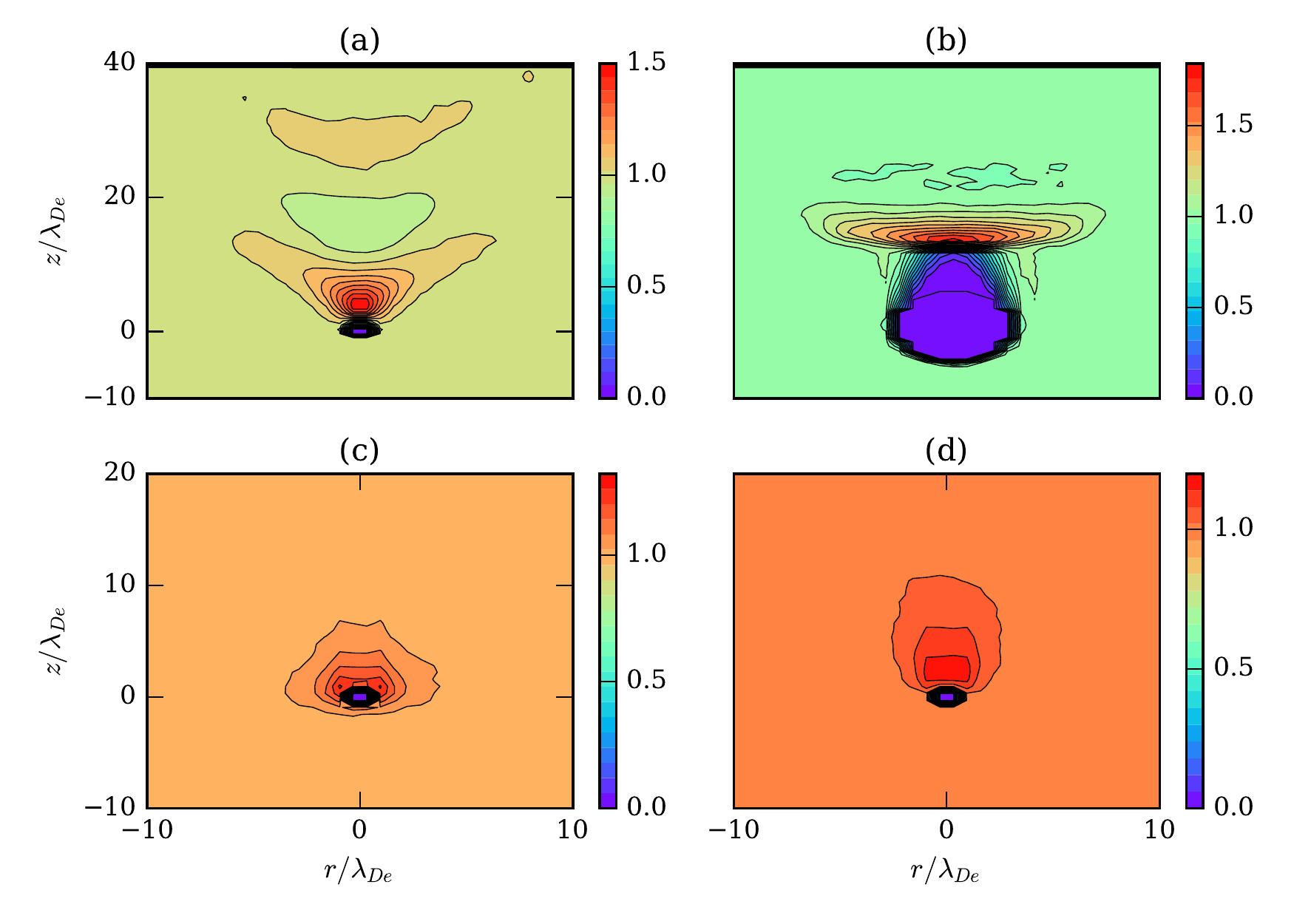} 
\caption{Spatial profiles of the ion density (scaled to the background), for a Mach number $M=0.8$, for a shifted Maxwellian (top row)
and a non-Maxwellian drift distribution (bottom row) for a collision frequency of $\nu =  0.01$ (left column) and  $\nu=0.1$ (right column).}
\label{fig:figure9}
\end{figure*}

In Fig.~\ref{fig:figure8}, we present the magnitude of the potential peak position and peak height as a function of Mach number $M$. The wake amplitude reveals a maximum with the Mach number~\citep{Ludwig:NJP2012}. The amplitude of the wake potential initially increases, with an increase of the streaming velocity. At moderate velocities it starts to saturate. Here, we also observe a non-monotonic behavior of the peak potential with the Mach number. The overall peak amplitude of the wake is higher, for the shifted Maxwellian distribution. One more feature we observe here is that the magnitude of peak position is higher for Maxwellian than in the non-Maxwellian case. 

The distinctions between the shift and drift distribution cases can also be elucidated clearly with the density profiles for the two cases. In Figure~\ref{fig:figure9} subplots (a) and (b), we present
the density profile for two different collision frequencies ($\nu = 0.01, \; 0.1$), in the Maxwellian case.  The density contours also show a wake pattern similar to the potential wake (for $\nu=0.01$). With increasing collisionality, the peak of the density decreases which is in agreement with the potential profiles. For lower collisionality, there is a large fraction of streaming ions that overcome the potential barrier and assist in
wake formation, whereas at higher collisionality the number of fast ion accumulation is reduced slightly. For $\nu=0.1$, the density bunching does not produce a wake pattern, rather there is big zero density region around the grain. The shielding of ions around the grain can be observed in the upstream direction as well. This is similar to the ``halo of ions shielding'' description given in Fig.4 of Ref.~\cite{Hutch:POP2013}.

In  subplots (c) and (d) of Figure~\ref{fig:figure9}, we present the density profile for two different collision frequencies ($\nu = 0.01, \and 0.1$) for the drift-driven non-Maxwellian distribution. Here, with increase in collisionality, the peak of the density increases, in contrast to the shifted Maxwellian case. In accordance to the corresponding potential profiles, here we do not observe ion density variations behind the grain. Moreover, slow ions create ion shielding around the grain, an effect that is missing for the shifted Maxwellian distribution~\cite{Hutch:POP2013}.

\section{Summary and Discussions}\label{sec:conclusion}
In the present work, we have investigated the electrostatic potential distribution around a point-like charged grain in a streaming plasma for a Maxwellian as well as for a non-Maxwellian ion distribution function. We have presented a comparative study of both cases, using accurate 3D particle-in-cell simulations.

One of the important observations is collision-induced amplification of the wake potential, for the realistic drift distribution case. While a shifted Maxwellian distribution also exhibits enhancement of the wake amplitude, for subsonic flows ($M\leq 0.3$), above this flow velocity range, ions are able to escape the potential well of the grain, and the dynamics are, altogether, different. For higher velocity values, the shifted Maxwellian distribution shows the damping of the wakefield with increasing collisionality. The basic difference between the two distributions is that the accumulation of ions behind the grain is completely different. The drift-driven non-Maxwellian distribution manifests collision-induced amplification, for the entire flow range considered in this paper. We have also presented the non-monotonic behavior with Mach number for the two distributions. The response of the grain wake features in the presence of collisions and  of an external field defies the common notion that an increase in collision frequency results in a damping of the  wake amplitude. We expect that our results will also be of direct relevance for experiments. Finally, an interesting question of current interest is how wake effects are modified in a magnetic field, where experiments~\cite{carstensen2012prl} and linear response simulations~\cite{Joost:PPCF2015,Hanno:CTPP2016} have observed a drastic modification of the wake potential with increasing field strength.

\section{Acknowledgments}
S. Sundar would like to thank I.H.~Hutchinson for support in using the COPTIC code, and acknowledges support of CAU Kiel. This work was supported by the DFG via SFB-TR24, project A9. Our numerical simulations were performed at the HPC cluster of Christian-Albrechts-Universit{\"a}t zu Kiel.

%\bibliography{dusty_manus1}
%

\end{document}